\documentstyle[aps,prb
]{revtex}

\begin{document}


\title{Monte Carlo simulations of infinitely dilute
solutions of amphiphilic diblock star copolymers}

\author{Ronan Connolly%
\setcounter{footnote}{1}\thanks{
E-mail: Ronan.Connolly@ucd.ie}}
\address{
Theory and Computation Group,
Department of Chemistry, University College Dublin,
Belfield, Dublin 4, Ireland}
\author{
Edward G.~Timoshenko%
\setcounter{footnote}{0}\thanks{Author to 
whom correspondence should be addressed. 
Internet: http://darkstar.ucd.ie; 
E-mail: Edward.Timoshenko@ucd.ie} 
}
\address{
Theory and Computation Group,
Centre for Synthesis and Chemical Biology,
Conway Institute of Biomolecular and Biomedical Research,
Department of Chemistry, University College Dublin,
Belfield, Dublin 4, Ireland}
\author{Yuri A. Kuznetsov%
\setcounter{footnote}{2}\thanks{E-mail: yuri@ucd.ie}}
\address{Centre for High Performance Computing Applications, 
University College Dublin,Belfield, Dublin 4, Ireland}

\date{\today}

\maketitle

\begin{abstract}
$\left.\right.$
\par\noindent
Single--chain Monte Carlo simulations of amphiphilic diblock star copolymers
were carried out in continuous space using implicit solvents. Two
distinct architectures were studied: stars with the hydrophobic blocks
attached to the core, and stars with the polar blocks attached to the core,
with all arms being of equal length.
The ratio of the lengths of the hydrophobic block to the length of the polar
block was varied from 0 to 1. Stars with 3, 6, 9 or 12 arms, each of length 10,
15, 25, 50, 75 and 100 Kuhn segments were analysed.
Four distinct types of conformations were observed for these systems. 
These, apart from studying the snapshots from the simulations, have been
quantitatively characterised in terms of the mean--squared
radii of gyration, mean--squared distances of monomers from the 
centre--of--mass, asphericity indices, static
scattering form factors in the Kratky
representation as well as the intra--chain monomer--monomer radial 
distribution functions. 
\end{abstract}

\section{Introduction}\label{sec:intro}

Due to recent advances in `living' polymer synthesis, by combining various
living polymerisation techniques such as anionic, cationic, ring--opening and
very recently, `pseudo--living' controlled radical polymerisations such as
atom transfer radical polymerisations, it is now possible to synthesise quite 
monodisperse star--branched polymers (i.e., a polymer with a single branch point,
the core, with all the branches, arms, of equal length), with a good degree of
control over the length and composition of the arms as well as the number of
them \cite{StarRev,StarCopolyRev,anionicStar,ATRPhomoStar,starROPandATRP}.  
Star--branched diblock copolymers, where each arm consists of an identical
diblock copolymer (as opposed to, for instance, heteroarm or miktoarm star
copolymers, where the arms themselves are of different
composition \cite{HeteroarmRev,HeteroarmA,HeteroarmB,HeteroarmC}) have been 
synthesised recently with increased
success \cite{CatAnionicStarDiblock,NonAmphStarDiblockA,NonAmphStarDiblockB,AmphStarDiblock1,AmphStarDiblock2,earlyInnHsyn,OutHsynA,OutHsynB,InnOutHsyn,OutHsynC,OutHUniInnHMix,InnHMix,SmidInnHf=3StarCMC,OutHf4Multi,InnHMix2}.
 
Of late there has been some interest in the potential use of amphiphilic
star copolymers
 \cite{earlyInnHsyn,OutHsynA,OutHsynB,InnOutHsyn,OutHsynC,OutHUniInnHMix,InnHMix,SmidInnHf=3StarCMC,OutHf4Multi,InnHMix2}
 (where one type of monomer is hydrophobic, H, and the other type is hydrophilic
or polar, P) as possible drug delivery vehicles. Star polymers have been shown
to have smaller hydrodynamic radii than linear
polymers \cite{OutHsynA,AdvantStarBlocks}, which could be important in avoiding
uptake by the reticuloendothelial system (RES) \cite{UhrichUni2}, as well as
having a lower crystallinity which is of importance in drug carrier
design \cite{AdvantStarBlocks}. It has also been suggested that star copolymers
could be designed which would form only unimolecular micelles in aqueous
solution \cite{OutHUniInnHMix,InnHMix,InnHMix2}. These systems would
have advantages as drug delivery vehicles over conventional multimolecular
copolymer micelles. Linear diblock copolymers have been found to form
multimolecular micelles above the critical micellar concentrations
(CMC) \cite{MicelleRev,AdvantPolymMicelle,PEOMicelleRev,MicelleDrugDelivA,MicelleDrugDelivB}.
These systems have been touted as potential drug delivery
vehicles, since they are able to encapsulate hydrophobic drugs. They can also be
designed to avoid uptake by the
RES \cite{OutHf4Multi,PEOMicelleRev,MicelleDrugDelivA,MicelleDrugDelivB}, since 
the polar blocks increase the `stealth' characteristics of the micelles. These
systems, however, are thermodynamically unstable below 
their CMC, and although polymeric micelles have been found to be more
thermodynamically stable than the simpler surfactant 
micelles \cite{AdvantPolymMicelle,PEOMicelleRev}, and they may be quite
kinetically stable, suggesting that they may be able to temporarily withstand
the  large dilutions involved in drug delivery \cite{MicelleRev,PEOMicelleRev},
these systems are still quite mobile, and drug molecules may still be able to
pass between the micelle and the
solution \cite{UhrichUni2,UhrichUni1,Arborescent}. 
It has been suggested that the thermodynamic instability of polymeric micelles
could be overcome by constructing branched polymers, where the amphiphilic
chains are covalently bound together \cite{UhrichUni2,UhrichUni1,Arborescent}. 

There appears to have been some success in creating these unimolecular micelles out of highly
branched systems, such as
dendrimers \cite{NewkomeDendUnimolecMicelle,MeijerDendOutHUnimolecMicelle},
graft copolymers \cite{UnimolecCombA,UnimolecCombB} and various other branched
topologies \cite{InnHMix2,UhrichUni2,UhrichUni1,Arborescent,FrechetUnimolecStarDend},
however the question of whether the single branch point of star copolymers is
sufficient to form unimolecular micelles, or whether they would form
multimolecular micelles, appears to be one of some debate at the moment, with
experimental evidence for both arguments. 
Regardless, it seems that some star copolymers are indeed capable of
forming purely unimolecular micelles, although
others form either a mixture of unimolecular micelles and multimolecular
micelles \cite{OutHUniInnHMix,InnHMix2}, or purely
multimolecular micelles \cite{InnHMix,OutHf4Multi,SmidInnHf=3StarCMC}. 

For instance, Miller {\it et al} synthesised different amphiphilic star
diblocks \cite{OutHUniInnHMix}. Using light scattering, they found that $f=6$
armed INNER--H and OUTER--H stars both are capable of unimolecular micelle
formation, although one of the systems also showed some limited aggregation. 
If chloroform is chosen, which is a good solvent for their
`H' monomers, and a poor solvent for their `P' monomers,
it appears that an INNER--solvophobic $f=6$ star, with
$N_{solvophobic}/N\approx$0.13 formed unimolecular micelles, with no evidence of
higher molecular weight aggregates. On the other hand, an OUTER--solvophobic
$f=6$ star, with $N_{solvophobic}/N\approx$0.32 did show some evidence of
aggregation, although predominantly unimolecular micelles were formed.

Gnanou {\it et al} synthesised two INNER--H, $f=6$ star diblocks \cite{InnHMix2}.
Analysis using size exclusion chromatography suggested that the stars with
$N_H/N$=0.03 formed unimolecular micelles in water. The stars with $N_H/N$=0.1
also formed unimolecular micelles, however aggregates of several stars were also
detected.
Interestingly, when they changed the solvent to tetrahydrofuran (THF) (of
opposite solvent quality for both monomer types),
in effect changing the stars to OUTER--solvophobic
topology, unimolecular micelles were still observed for the
$N_{solvophobic}/N$=0.9 star (i.e., the $N_H/N$=0.1 star). Of course,
aggregation was more dominant than in water, however a significant proportion of
the conformations were suggested to be unimolecular in nature. In contrast, the
$N_{solvophobic}/N$=0.97 star (i.e., the $N_H/N$=0.03 star) showed little
tendency for unimolecular micelle formation, favouring intermolecular
association.

Several groups were unable to detect unimolecular micelle formation for their
synthesised amphiphilic star diblocks, instead detecting quite distinct
CMC \cite{InnHMix,SmidInnHf=3StarCMC,OutHf4Multi}. However these studies all
involved $f=3$ or $4$ armed stars, and it is possible that more arms are indeed
necessary for unimolecular micelle formation, as suggested earlier.

Uhrich's group have synthesised novel amphiphilic polymer systems, which have
some similarities with $f=3$ star diblocks \cite{UhrichUni2,UhrichUni1}. They
find that these polymers appear to form unimolecular micelles exclusively in
aqueous solution. Although their polymers contain only $f=3$ polar,
poly(ethylene glycol), arms, 
they suggest that these arms are not fully extended, but instead wrap
around the hydrophobic core, presumably offering enough `shielding' to prevent
intermolecular aggregation.

It seems likely that the CMC is at least significantly reduced, 
when present, for star diblock copolymers in comparison to linear 
diblock copolymers, since several of the diblocks
are already covalently joined through the core monomer. 
There is also reason to suggest that these systems even have micellar behaviour
(potentially allowing encapsulation) below their CMC \cite{SmidInnHf=3StarCMC}.

The single--chain behaviour of these systems seems particularly
important as it serves as a prerequisite for trying to understand the
effects of higher concentrations.  Computer simulations have
been really successful in investigating single linear chains, but
there does not appear to be many computational
studies of isolated amphiphilic star diblock copolymers. 
Aside from a previous work involving two of the coauthors
of the current paper \cite{copstarFabio}, the only other appropriate 
simulational study we could find was a Monte Carlo study by Nelson {\it et
al} \cite{Nelson}. This work had some limitations which we have previously
discussed \cite{copstarFabio}, in particular, the fact that it was performed using a
lattice. As we have previously mentioned \cite{copstarFabio}, lattice simulations
can be problematic for branched systems. We believe that for them
the additional computational expense required to run Monte Carlo
simulations in continuous space is well justified and, in fact, 
quite important.

In our previous study of star copolymers \cite{copstarFabio}, we only briefly 
discussed the effect of varying the hydrophobic:polar (H:P) composition of 
the arms, with the bulk of the study being dedicated to the 50:50 cases. 
In reality, however, there appears to be quite a range of H:P ratios
synthesised \cite{earlyInnHsyn,OutHsynA,OutHsynB,InnOutHsyn,OutHsynC,OutHUniInnHMix,InnHMix,SmidInnHf=3StarCMC,OutHf4Multi,InnHMix2}.
Control of the H:P ratio is particularly significant with regard to drug 
delivery applications. Low values of H:P seem to increase the 
stealth characteristics in evading uptake by the RES \cite{MicelleDrugDelivA,MicelleDrugDelivB},
as well as being more likely to result in the formation of unimolecular
micelles \cite{InnHMix2}. 
On the other hand, high values of H:P tend to increase the drug encapsulation
ability of micelles \cite{MicelleRev}, while reducing the drug release
rate \cite{MicelleDrugDelivA}, as well as reducing the CMC of those systems
which form multimolecular micelles \cite{MicelleRev}. Hence we wish to address
this issue in more detail here.

Moreover, we have previously seen in
Ref. \onlinecite{copstarFabio} that an additional metastable state with
compacted H subglobules at the ends of outstretched P arms occurs for
the OUTER--H star copolymers. Here we would like to investigate
whether this state could indeed become thermodynamically stable for
some range of compositions as conjectured, as well as to look at the 
phase diagram of such stars more generally. 
Also, we would be interested to study several useful
observables of star copolymers, most of which are 
experimentally measurable,  such as the 
asphericity characteristics, static structure factors (SSF), and 
intra--chain monomer--monomer radial  distribution functions (RDF), 
which have been recently investigated by us for the homopolymer stars and 
dendrimers in the good solvent \cite{dend,corrfn}.


\section{Model and notations}
\label{sec:model}

The current coarse--grained star copolymer model is based
on the following Hamiltonian (energy functional) 
 \cite{copstarFabio,dend,corrfn} 
in terms of the monomer coordinates, ${\bf X}_i$:
\begin{eqnarray}
\frac{H}{k_B T} & = & \frac{1}{2\ell^2}  \sum_{i\sim j} \kappa_{ij}
        ({\bf X}_i - {\bf X}_{j})^2
+\frac{1}{2} \sum_{ij,\ i\not= j}
U^{(lj)}_{ij} (|{\bf X}_i - {\bf X}_j|).
\label{cmc:hamil}
\end{eqnarray}
Here the first term represents the connectivity of the star with
harmonic springs of strength $\kappa_{ij}$ introduced between
any unique pair of connected monomers (which is denoted by $i\sim j$). 
It should be noted that we define a `monomer' in relation to 
the Kuhn length of the
polymer, and hence more detailed bonded terms are not essential here.
The second term represents pair--wise non--bonded, viz., van der Waals,
interactions between monomers.
We shall adopt the Lennard--Jones form of the potential,
\begin{equation}\label{VLJ}
U^{(lj)}_{ij}(r) = \left\{
\begin{array}{ll}
+\infty, &  r < d^{(0)}\\
U^{(0)}_{ij} \left( \left( \frac{d^{(0)}}{r}\right)^{12}
- \left( \frac{d^{(0)}}{r} \right)^{6} \right), &
r > d^{(0)}
\end{array}
\right., \qquad
U^{(0)}_{ij} = \frac{1}{2}\left(  U^{(0)}_{i} + U^{(0)}_{j}\right).
\end{equation}
with a hard core part and the monomer diameter $d^{(0)}$
(equal for H and P monomers in this work),
which will be chosen to coincide with $\ell$ here,
where $U^{(0)}_i$ is the dimensionless strength of the 
Lennard--Jones interaction acting upon monomer $i$. 

In amphiphilic copolymers, there are two types of monomers, hydrophobic (H) and
polar (P). We model the behaviour of these two types in aqueous solution with
the following method incorporating the effect of the solvent into
`effective' monomer--monomer interactions. 
We define the strength of the interactions between two hydrophobic beads as
$U^{(0)}_{HH}=5\, k_B T$ here. The strength of the interactions between two 
polar beads is defined as $U^{(0)}_{PP}=0$. 
We take the strength of the interactions between a hydrophobic bead and a polar
bead as being the arithmetic mean of the two possible interactions, i.e.,
$U^{(0)}_{HP}=2.5\, k_B T$. This implicit solvent model and
particular choice of parameters, while physically simple, 
are somewhat non--trivial to derive, but they have been previously well
rationalised \cite{copstarFabio,YuriCopoly}.

Strictly speaking, this model is generic in the description of the solvent, and
so we could think of the monomers as solvophobic and solvophilic, but since
much of the work on star diblock copolymers is interested in aqueous solution
and in order to allow easy comparison with our earlier study of these systems
\cite{copstarFabio}, it is convenient to use the hydrophobic/polar notation.

We will also use the other notations from our earlier study where appropriate: 
Each star consists of $f$ arms, or functionalities, attached to a single core
monomer. The total number of monomers is defined as $N+1$, resulting in $N/f$
monomers in each arm, assuming that all arms are of the same length.
For diblock stars, consisting of a block of H beads and a block of P beads, we
will refer to the situation where the H-block is attached to the core monomer
and the P-block is on the outside as `INNER--H', as in our previous
study \cite{copstarFabio}. 
We will refer to the reverse situation, where the P-block is attached to the
core monomer and the H-block is on the outside as `OUTER--H'. 
In both cases, for simplicity, we assume that the core monomer has the
properties (i.e., H or P) of the monomers to which it is attached. 
The composition of the arms will be denoted by the parameter, $N_H/N$ or
$N_P/N$, where $N_H$ is the number of hydrophobic beads and $N_P$ is the number
of polar beads. 

Now let us introduce the main observables of interest.
The mean--squared (MS) radius of gyration and the MS distance of monomer $i$
from the centre of mass, are defined respectively as,
\begin{eqnarray}\label{Rg}
\langle R_g^2 \rangle = \frac{1}{2N^2}
               \sum_{ij,\ i\not= j} \biggl\langle ({\bf X}_i - {\bf X}_{j})^2 \biggr\rangle,
\qquad 
\langle R_{i}^2 \rangle = \biggl\langle ({\bf X}_i - \frac{1}{N} \sum_{j} {\bf
X}_j)^2 \biggr\rangle. 
\end{eqnarray}
The asphericity and ellipsoid indices, $A_3$ and $S_3$ respectively, are defined
as in Ref. \onlinecite{corrfn}, such that 
\begin{eqnarray}
\qquad 0\ \mbox{(sphere)} \leq {\cal A}_3 \leq 1
\ \mbox{(collinear)}, \label{Aa}\qquad
 -\frac{1}{4} \ \mbox{(oblate)} \leq {\cal S}_3
\leq 2 \ \mbox{(prolate)}. \label{Ss}
\end{eqnarray}
The dimensionless specific heat capacity is related to the energy variance,
\begin{equation}
\frac{c_V}{k_B}=\frac{\langle(\Delta E)^2 \rangle}{N(k_BT)^2}.
\end{equation}

The intra--chain monomer--monomer radial distribution function (RDF) of a pair
of monomers $i$ and $j$ is defined as,
\begin{equation}
g^{(2)}_{ij}({\bf r}) \equiv \biggl\langle \delta({\bf X}_i
- {\bf X}_{j}-{\bf r}) \biggr\rangle = \frac{1}{4\pi r^2} \biggl\langle
\delta(|{\bf X}_i - {\bf X}_{j}|-r) \biggr\rangle.
\end{equation}
as in Ref. \onlinecite{corrfn}.
The function is normalised to unity via:
$\int d^3{\bf r}\, g^{(2)}_{ij}({\bf r})=1$.
The static structure factor (SSF), is introduced as follows:
\begin{eqnarray}\label{eqn:S(q)}
S(q) &=& \frac{1}{N}\sum_{ij} \tilde{g}^{(2)}(|{\bf q}|), \qquad
\tilde{g}^{(2)}({\bf q}) = \left\langle \exp(i{\bf q}\,({\bf X}_i -
    {\bf X}_j) \right\rangle = \frac{1}{2\pi^2}
    \int_0^{\infty}r^2\,dr\,\frac{\sin(qr)}{qr}\,g^{(2)}_{ij}(r),
\end{eqnarray}
where tilde indicates the 3D Fourier transform, and the wave number
is $q=4\pi\sin \left(\frac{\theta}{2} \right)/\lambda$.
Partial SSFs are calculated using a similar definition, except
that summation only occurs over the relevant monomers.

\section{Simulation Techniques}
\label{subsec:cmc}

We use the Monte Carlo (MC) technique with the standard Metropolis 
algorithm  \cite{Metropolis,AllenTild} and local monomer moves 
based upon the implementation 
described by us in Refs.~\onlinecite{copstarFabio,dend,corrfn} and references 
described therein. 
A Monte Carlo sweep (MCS) is defined as $N$ attempted MC steps.
Initial conformations of stars were taken as globules obtained from simulations
of the equivalent homopolymers in poor solvent.
This was done in order to reduce computational expense and avoid 
possible metastable states which were observed in a previous study \cite{copstarFabio}, as
it takes longer for two H beads to meet than for two P beads to separate.
These conformations were however then subjected to extensive equilibration for a
required time before any simulation was commenced.

To ensure good equilibration, the behaviour of global observables such as the
energy and radius of gyration were monitored.  
After reaching equilibrium, a large number of statistical measurements were
performed. To ensure statistical independence of sampling, each consecutive
measurement was separated by a large number of MCS. The number of MCS required
between each measurement, was calculated by ensuring that the `statistical
inefficiency', s, of relevant observables tended towards 1, as described in
Ref. \onlinecite{AllenTild}.
The mean value and error of sampling of an observable $A$ are then
given by the arithmetic mean 
$\langle A \rangle=(1/Q)\sum_{\gamma}^Q A_{\gamma}$
and by $\pm \sqrt{(\Delta A)^2/Q}$ respectively,
where Q is the number of statistically independent measurements
\cite{footnote}. As in our previous works 
\cite{copstarFabio,corrfn,dend}, energies will be reported in
units of $k_BT$ and distances will be reported in units of $\ell$.


\section{Results from Monte Carlo simulations}
\label{subsec:molres}

We have studied diblock star copolymers with $f=3,\ 6,\ 9$ 
and $12$ arms, each of 
length $N/f=10,\ 15,\ 25,\ 50,\ 75,\ 100$ monomers. 
The relative ratio of the length of the P-block to the total arm length
was varied from $N_P/N$=0 to 1. Obviously, since the arms were of finite length,
only certain discrete values of this ratio were permitted. 
The two distinct INNER--H and OUTER--H topologies
described in Section \ref{sec:model} were investigated.

\subsection{Possible conformations and phase diagram}
\label{subsec:phd}

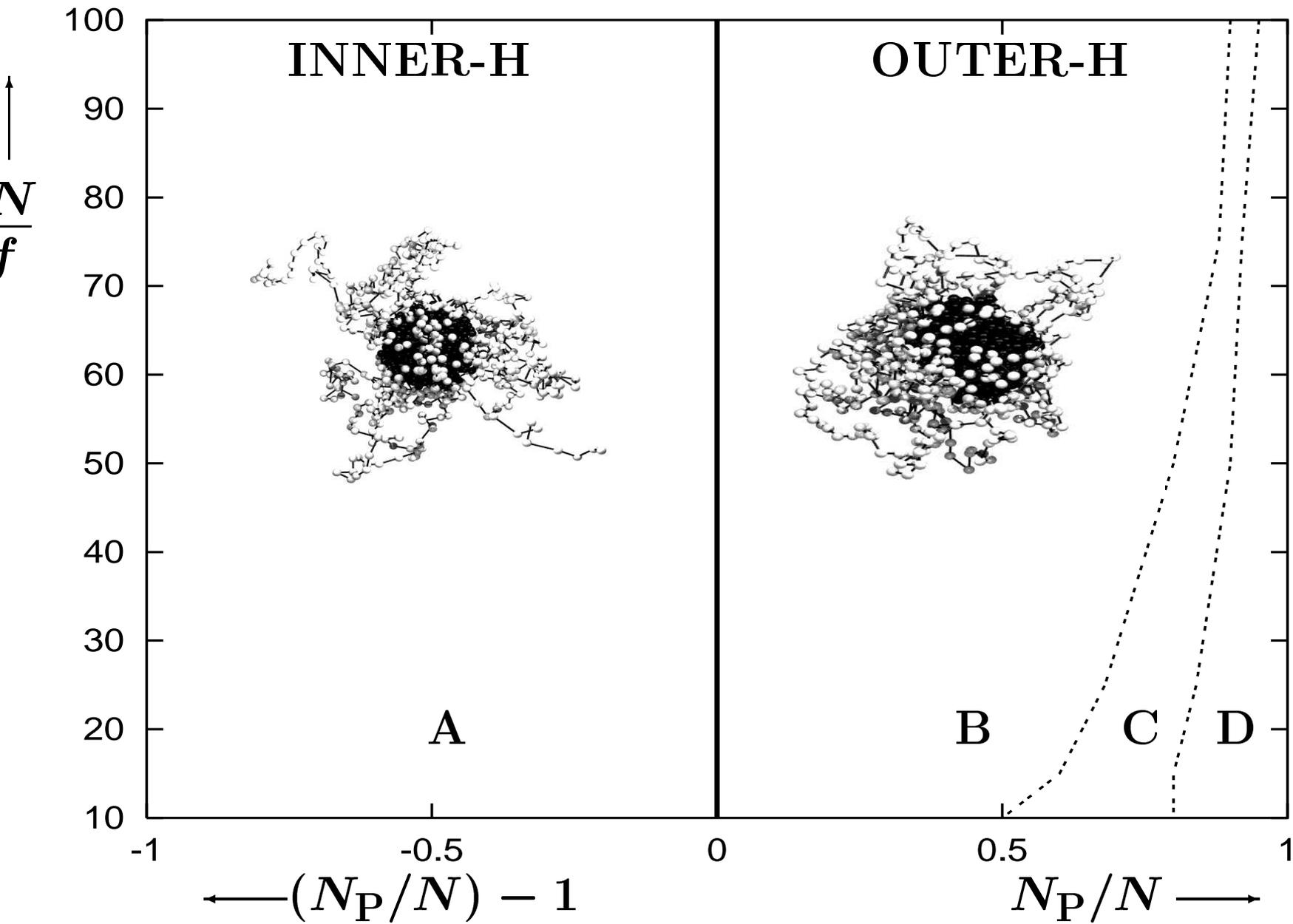
\begin{figure}
\caption{
\label{fig:phase}
A phase diagram of star copolymers with a fixed
number of arms $f$ obtained from the data
for stars of $f=3,6,9,12$ arms with
illustrations of their conformations in terms of the
composition, $N_P/N$, and the arm length, $N/f$. 
Both INNER--H and OUTER--H stars are represented on the same diagram
by showing the composition of INNER--H
stars with $(N_P/N)-1$. Thus, the leftmost (rightmost) points on 
both parts correspond to totally H (P) homopolymer stars.
Distinct possible conformations are designated by the
letters A, B, C and D. Typical snapshots for A and B are also incorporated
into the figure (these are shown for stars with f=12 arms of
100 monomers each). Snapshots for C and D are shown in Fig. \ref{fig:snapsC+D}.
Here and below the light gray spheres represent polar (P) monomers, 
while the black spheres represent hydrophobic (H) monomers. 
}
\end{figure}

Four distinct types of conformations were observed for these systems, 
labeled A, B, C and D in Fig. \ref{fig:phase}. 
Types A and B correspond to those previously observed in Ref. \onlinecite{copstarFabio}.
Namely,
for INNER--H stars, the H beads cluster together to form a globule, while 
the monomers in the P-blocks behave as coils, albeit attached to 
the globule, resulting in the usual micellar conformations (type A). 
For most OUTER--H stars a micellar type of conformation is also observed, 
with the H beads forming a globule and the P-blocks having coil--like behaviour, 
although in this case, 
due to connectivity constraints, the P-blocks are all attached to the 
polar core monomer located outside the H globule,
as well as to the H globule itself. 
Such a conformation (type B) can hence be described as a micelle with 
daisy--like loops, or loopy micelle in brief. 
It can be seen that the extra attachment of the P-blocks necessary
for type B is entropically unfavourable, although 
energetically favourable as it permits the H beads to form
a compact globule, resulting in an overall lowering of the Helmholtz
free energy. 
However, if the number of H beads is decreased, corresponding to 
larger values of $N_P/N$, then
this lowering in energy is reduced and may not be enough to counteract the loss
in entropy required for the H beads at the ends of each P-block to meet. 
Hence for large values of $N_P/N$ another type of conformation (type D) is
observed, where the H-beads at the end of each arm form separate globules.
Here, however, these globules do not coalesce, and so the observed 
conformations are akin to those of star homopolymers in good solvent, 
although with $f$ clusters of H beads at the ends of the arms.  
A typical snapshot with this type of conformation is illustrated in Fig.
\ref{fig:snapsC+D}b. Although such a shape corresponded to a
metastable conformations for compositions considered 
in Ref. \onlinecite{copstarFabio}, 
these conformations, in the area of the phase diagram designated as D, 
correspond to the true equilibrium structures. These never change
during the equilibrium simulation no matter how long and, importantly,
do not depend on the initial conformation or the pathway leading
to the part D of the phase diagram, reaffirming that the state is
indeed an equilibrium one.

For slightly smaller values of $N_P/N$, an intermediate regime is
observed, where some of these globular clusters can coalesce, but not all of
them together. We refer to these types of conformations as type C. 
In this area of the phase diagram, C, states with such partial clustering
have a lower free energy than any of the states B or D, and hence the
area C also corresponds to equilibrium structures, related
to different ways of clustering in its different parts, and not merely
local free energy minima. However, interestingly,
some of these globular clusters
can enter and leave the larger globular subglobules 
throughout the course of the simulation, indicating that the barriers 
between such distinct minima are fairly low.
For instance, a typical snapshot of type C is illustrated in Fig.
\ref{fig:snapsC+D}a.  
Although there are three clusters of H beads (black) in this snapshot, the
actual number of clusters varied from three to five in the course of the
simulation. An analysis of the energy suggests that any energy barriers 
between these different structures are easily traversed by the thermal 
fluctuations present in the simulation, although the entropy difference
between such states is hard to quantify from the Monte Carlo data.
The resulting cluster size distribution is a fairly broad one,
as opposed to the fairly narrow aggregation size distributions seen
previously for the `mesoglobules', which represent multimolecular
self--assemblies of linear heteropolymers of certain sequences
in narrow strips of the phase diagram \cite{YuriCopoly}.
 
A comparison of the heat capacity values presented in
Fig. \ref{fig:C_V} for stars with $f=6$ arms of length $N/f=50$ beads reveals
its relatively high values for the systems which adopt conformations of type C,
i.e., those where $N_H/N=0.1,\, 0.12$ and $0.14$. This is indicative of greater
fluctuations in energy associated with
several differently clustered conformations with low enough energy 
barriers between the corresponding energy minima that are
thermally accessible to the system.
Increasing the length of the arms
reduces the entropic loss required for conformations of type B, and so the range
of values of $N_P/N$ with conformations type B is increased. 
Changing the number of arms with a fixed arm length, $N/f$, did
not seem to have any significant effect on the phase diagram, although stars
with more than $f=12$ arms were not studied in this work. Therefore,
we could extrapolate the phase boundaries between B, C and D regions 
using the data available for stars with different number of arms
$f=3,\,6,\,9$ and $12$.

With these four types of conformations in mind, it is possible to 
explain the different values of
MS radii of gyration, $\langle R_g^2 \rangle$, 
which can be used as an
indicator of the size of the polymer plotted in Fig.
\ref{fig:R_g^2} vs $N_H/N$ for INNER--H and OUTER--H stars. In
Fig. \ref{fig:R_g^2}, we only discuss $\langle R_g^2 \rangle$ for stars with
$f=6$ arms of length $N/f=50$ beads, 
however the other systems studied were qualitatively
similar. For INNER--H stars there is a relatively uniform decrease in $\langle
R_g^2 \rangle$ with increasing $N_H/N$, since the compact globular component of
the micelles increasingly dominates the size as the relative fraction of the
H beads is increased.  
A similar behaviour is observed for the OUTER--H stars with $N_H/N>0.14$,
corresponding to conformations of type B (loopy micelles) for the same reason.
Clearly, it can be seen that the $\langle R_g^2
\rangle$ values are considerably smaller for the OUTER--H stars, since 
the size of type B conformations is more
compact due to the additional tethering of the P-blocks to the core monomer. 
However, for stars with $N_H/N<0.1$, the 
OUTER--H stars in fact have higher $\langle
R_g^2 \rangle$ values than their corresponding INNER--H stars. 
This results from the fact that the type D conformations observed for these stars 
are less compact
than the type A conformations, since the H beads form several clusters at the
end of each outstretched arms (See Fig. \ref{fig:snapsC+D}b),
instead of the single H bead globule observed for type A stars (See Fig.
\ref{fig:phase}).
As mentioned before, the OUTER--H stars with $N_H/N=0.1,\ 0.12$ and $0.14$ have
conformations of type C, an area where a rapid fall of the 
$\langle R_g^2 \rangle$ occurs. 
Obviously, the stars with $N_H/N=0$ and 1 correspond to
the homopolymers in good and poor solvents respectively, and hence are identical
in size for `INNER--H' and `OUTER--H'. 

In order to obtain a more detailed understanding of the various types of conformations which occur
for OUTER--H stars, it may be helpful to compare the MS distances plots of each bead, $i$,
from their centre-of-mass for various values of $N_P/N$, shown in Fig.
\ref{fig:R_i^2}. Fig. 17 in Ref. \onlinecite{copstarFabio} illustrates these plots for the INNER--H case,
and while Fig.~18 in Ref. \onlinecite{copstarFabio} illustrates the OUTER--H case, the plots for
values of $N_P/N>0.8$ are not reported there. 
For the stars discussed here it is those values of
$N_P/N>0.8$ which correspond to types C and D, so it is useful to explicitly include 
the plots for the region $N_P/N=0.8-1.0$. 
The plots discussed here are again those corresponding to stars with
$f=6$ arms, each of length $N/f=50$ beads. For the stars with $N_P/N<0.86$, corresponding to type
B (lower curves in the middle of the Fig.), 
the beads have a maximum value for the bead in the middle of the P-block, since the other ends
are attached to either the H-globule or the core bead, as has been discussed previously in 
Ref. \onlinecite{copstarFabio}. The MS distances of the H beads from the centre of mass are 
fairly constant  and small, as is characteristic of the globules, whereas 
those of the P blocks are much larger and have a distorted bell shape.
The stars with $N_P/N>0.9$ adopt conformations of type D
(upper curves in the middle of the Fig.). 
These stars are only tethered at the core
bead, hence the MS distances from the centre-of-mass, 
which in this case correspond to distances near the core
bead, increase along the arm, until a plateau is reached for the H beads, 
as these are all in a subglobule and hence close to each other spatially. 
Intermediate values of $N_P/N$ correspond to
conformations of type C. As was already mentioned stars of this type have 
conformations with variable degrees of clustering. 
Stars of type C, which on average have larger numbers of H clusters tend to be more
similar to type D than type B, e.g., $N_P/N=0.9$, while those which on average 
have smaller numbers of H clusters, tend to be more similar to type B than 
type D, e.g., $N_P/N=0.86$ and $0.88$.

\subsection{Asphericity characteristics}

Increasing the number of arms in homopolymer stars is known to cause more
spherical conformations \cite{corrfn,StarAspher}. This also occurs for the
copolymer stars discussed here, however there does not appear to be any other
effect of increasing the number of arms in the range studied here, so we will
limit our discussion here to the case of $f=6$ arms. 
The following discussion can also be applied to the cases of $f=3,\ 9$ or $12$
arms,
except that the absolute values of $A_3$ and $S_3$ decrease with increasing
numbers of arms.

In Figs. \ref{fig:A_3} and \ref{fig:S_3} the asphericity and ellipsoid indices 
respectively of both INNER--H and OUTER--H stars with $f=6$ arms each of length
$N/f=50$ beads are plotted against the H composition, $N_H/N$. 
It can be seen that OUTER--H stars of type B ($N_H/N>0.14$) are more spherical
(hence lower $A_3$) and less prolate (hence lower $S_3$) than their type A 
INNER--H counterparts. This is again due to that type B stars are more compact,
even though they have certain asymmetry related to 
the fact that the core bead is outside the H subglobule. 
However, for low values of $N_H/N$, i.e., $N_H/N<0.08$, there is very little
difference between the INNER--H and OUTER--H stars in terms of asphericity and
prolateness, since the end monomers contribute little to these characteristics,
in either types A or D. Generally,
for both types A and B, increasing $N_H/N$ results in more spherical 
shapes as the H subglobule increasingly dominates. 
As can be observed in Fig. \ref{fig:snapsC+D}a, the typical snapshots of type C
stars are quite asymmetrical, explaining the relatively large values for both
$A_3$ and $S_3$ in the region corresponding to type C stars.

\subsection{Intrachain monomer--monomer radial distribution functions}
\label{subsec:gr}

The monomer--monomer radial distribution functions (RDF), 
$g^{(2)}_{ij}(r)$, between
monomers $i$ and $j$, are useful in providing statistical information on the
internal structure of conformations adopted by polymers \cite{corrfn}.
We report here the rescaled versions of these functions, in terms of
dimensionless variables after rescaling via the MS distances between monomers
$i$ and $j$, as discussed
in Ref.  \onlinecite{corrfn}, namely,
\begin{equation}
\label{gHat}
\hat{g}_{ij}^{(2)}(\hat{r}_{ij})\equiv  D_{ij}^{3/2} g_{ij}^{(2)}(r),
\qquad \hat{r}_{ij}\equiv \frac{r}{\sqrt{D_{ij}}},
\qquad D_{ij} \equiv \langle ({\bf X}_i - {\bf X}_j)^2 \rangle.
\end{equation}
This rescaling allows one to perform a more straightforward comparisons
between different systems and sizes and highlights the universality
properties of the RDF \cite{corrfn}.

Since there are large numbers of different $\hat{g}_{ij}^{(2)}$
types, we would be only interested in discussing here the novel features
present in copolymer stars. 
In the poor solvent, the topology of most flexible homopolymers becomes less
significant, except locally \cite{dend}.
Indeed, the H-H RDFs behave pretty much as 
in the homopolymer globules with an oscillating shape.
Behaviour of most of the P-P RDFs are also broadly akin to those of 
homopolymer coils with a power law fall at small distances (`correlation hole') and a stretched
exponential decay at large ones \cite{corrfn}, although the effect of
double constraint in the loopy micelle (type B) is quite significant.
Nevertheless, it would be most interesting to investigate the mixed H-P RDFs
which have no homopolymer analogues.
As the monomers inside the H globules are nearly equivalent to
each other as we have seen above, we can further average
the functions $\hat{g}_{ij}^{(2)}$ over $i$ values belonging to the H section,
thereby improving the statistics.

From Figs. \ref{fig:g_rArms} and \ref{fig:g_rOutH}, we can determine how
certain representative monomers in the polar block align themselves with regard
to the hydrophobic globular blocks, in various different stars. 
We report these plots for $N_H/N=N_P/N=0.5$ composition. 
Na\"\i vely, one could expect that the hydrophobic globule can simply be
treated as a `core monomer', albeit a rather large one. Hence the
$\hat{g}^{(2)}_{H,end}(\hat{r})$ functions should behave similarly to the
$\hat{g}^{(2)}_{ij}(\hat{r})$ functions between monomers in the arms and the
core monomer of a homopolymer star in good solvent, as in Ref.  \onlinecite{corrfn}. 
Indeed, if one neglects the interactions between hydrophobic monomers and polar
monomers, this is what would result. 
As we have mentioned previously in Section \ref{sec:model}, we treat the
H-P  interactions as being the arithmetic mean of the interactions between
two monomers of the same type, using the rationale discussed in Refs.
\onlinecite{copstarFabio,YuriCopoly}.
This model then yields a slight attraction of polar monomers for the hydrophobic
monomers in the globule for the parameter values considered here.
This weak very short--ranged attraction causes the polar blocks to
wrap around the hydrophobic globule,
resulting in the occurrence of a peak for small values of
$\hat{r}$ in $\hat{g}^{(2)}_{H,end}(\hat{r})$ near the steric contact 
separation for the RDF between the end P-monomers
and the H-monomers of INNER--H stars with $f=3,6,12$ arms in
Fig. \ref{fig:g_rArms}. Other small
peaks (reflecting the `microstructure' of the conformation) 
correspond to higher order preferential solvation
`shells' of the H-subglobule as in Fig. 5 in Ref. \onlinecite{corrfn}.
These peaks become less significant with increasing number of arms. 
However, the behaviour for large values of $\hat{r}$ is quite important
as the tails of the plots with increasing numbers of arms do not overlap 
for large values of $\hat{r}$ in Fig. \ref{fig:g_rArms}. 
This can be explained by the increased steric repulsion near 
the centre of the star with larger $f$, making it more
favourable for polar monomers to adopt positions away from the core. 
This is also manifested in the decreasing values of 
$\hat{g}^{(2)}_{H,end}(\hat{r})$ for small
$\hat{r}$, since increasing $f$ promotes
the `correlation hole' effect leading to higher values
of the power--law exponent $\theta$ (see second block in Tab. II
of Ref. \onlinecite{corrfn}).

It is also interesting to compare the RDFs of the INNER--H and OUTER--H 
cases. In Fig. \ref{fig:g_rOutH}a, we compare the
$\hat{g}^{(2)}_{ij}(\hat{r})$ plots between the hydrophobic globule and the
monomer at the end of the polar block (i.e., the end monomers for INNER--H
topologies and the core monomer for OUTER--H topologies), for both types of
stars, with the same number of arms. The RDF 
$\hat{g}^{(2)}_{H,end}(\hat{r})$ of INNER-H is more localised than
$\hat{g}^{(2)}_{H,core}(\hat{r})$ of OUTER-H stars in this rescaled
representation, while we remember that 
the corresponding $D_{ij}$ of OUTER-H stars is actually considerably smaller
than for INNER-H. 
This can be easily explained according to
our rationale in Ref. \onlinecite{corrfn}. Indeed, 
the large-$\hat{r}$ stretching exponent,
$\delta=1/(1-\nu)$, for the INNER-H case gives a value
close to $2.5$ as $\nu\simeq 3/5$ for the end-H interactions as in the
homopolymer P-star. On the other hand, the highly constrained P-blocks
of the OUTER-H loopy micelle have nearly quasi--ideal behaviour, and
hence a more compact size, with 
$\nu\simeq 1/2$ yielding $\delta \simeq 2$ for the core-H interactions.
We also see the first solvation and higher
order microstructure peaks present here as well.

Fig. \ref{fig:g_rOutH}b compares the behaviour of two different monomers in
the polar block of the same OUTER--H star. It can be seen that the RDF
$\hat{g}^{(2)}_{H,mid}(\hat{r})$ decreases somewhat faster with $\hat{r}$ 
than $\hat{g}^{(2)}_{H,core}(\hat{r})$ for these $f=12$ arms stars.
This subtle effect can be also interpreted via a larger stretching
exponent $\delta$ for the middle-H correlations as the middle
monomers close to the H-block do not quite reach the Gaussian behaviour
due to higher stretching of respective P-segments.
Again, we also see by comparing the functions $\hat{g}^{(2)}_{H,core}(\hat{r})$ 
in Figs.~\ref{fig:g_rOutH} a and b
that the microstructure oscillations decrease with increasing number of arms.

\subsection{Static structure factors}

Although the static structure factors (SSF) 
are somewhat less informative than the monomer-monomer radial
distribution functions \cite{corrfn}, SSF can be
readily obtained from light and neutron scattering techniques
\cite{corrfn,NeutScattStar,SelectiveNeutScattStar}. 
In particular with neutron scattering it is possible to obtain partial form
factors, due to the fact that by hydrogen--deuterium isotope exchange the
scattering contrast can be changed at will, allowing one to observe structural
properties of selected parts of the star \cite{SelectiveNeutScattStar}. 
It is convenient to report SSF using rescaled Kratky forms as follows, 
\begin{equation}
\hat{S}(\hat{q})=\frac{q^2R_g^2}{N} S(q)\textrm{, }\qquad
\hat{q}=q\sqrt{R_g^2}\textrm{, }
\end{equation}
since this is less sensitive to effects of varying N. 
The Kratky representation is convenient for providing information about the
conformations adopted by polymers, as it highlights the deviation of the form
factor of the polymer from that of a linear Gaussian coil for large values of
$q$, which would reach a constant asymptote.
Note that for the partial SSF the respective partial $R_g^2$ is used
for rescaling.

In Figs. \ref{fig:SFFInnH} and \ref{fig:SFFOutH}, we report the SSFs
calculated from the simulation results obtained for stars with $f=6$ arms, each of
length $N/f=50$ beads, and INNER--H and OUTER--H topologies respectively. The
cases where the H and P blocks are of 
equal length, i.e., $N_H/N=0.5$, are reported here. The relative statistical
errors of these plots are smaller than the resolution can distinguish, due to the
extensive averaging inherent in the function, and so are not included.

We have discussed in a previous work \cite{corrfn} the SSFs of homopolymer stars
in good solvent with increasing arm numbers as well as several other systems.
Freire {\it et al} have also calculated SSFs for several homopolymer stars in
good solvent, using results from lattice Monte Carlo
simulations \cite{StarKratky}. 
As found experimentally \cite{NeutScattStar,SelectiveNeutScattStar}, for small
$\hat{q}$ the SSFs of most systems are similar when using the Kratky forms.
Similar behaviour is observed in Figs. \ref{fig:SFFInnH} and \ref{fig:SFFOutH}
for the universal area $\hat{q}< 1$. 
For linear polymers in poor solvent, i.e., globules, a
maximum in $\hat{S}(\hat{q})$ is typically reached shortly
afterwards \cite{corrfn}.  $\hat{S}(\hat{q})$ then decreases until it reaches a
value near zero, where it begins oscillating with
increasing $\hat{q}$. This is indicative of the dense structure of a globule.
$\hat{S}(\hat{q})$ for linear polymers in good solvent, i.e., coils, on the other
hand, continues to increase for increasing $\hat{q}$ \cite{corrfn}, indicative of
the loose structure of coils. Stars in good solvent tend to be denser than
linear polymers, and so they adopt an intermediate behaviour, with
$\hat{S}(\hat{q})$ initially decreasing after an initial maximum as for the
dense globules, before starting to increase as for coils. This deviation from
the behaviour of linear polymers in good solvent becomes more pronounced for
larger numbers of arms \cite{corrfn,StarKratky}, corresponding to the more dense
nature of many--arms stars. 
In poor solvent, the connectivity of stars is less significant, and they
behave similarly to linear polymer globules. Na\"{\i}vely, one could assume that
diblock star copolymers, where half of the beads are in good solvent and half of
the beads are in poor solvent, would have an intermediate behaviour between the
two homopolymers. It should be remembered, however, that the behaviour of beads in
the inner region of the star is different from the outer region of the star even
for the simpler homopolymer in good solvent \cite{SelectiveNeutScattStar}. The
inner region of stars tends to be more densely packed than the outer region. For
this reason, as well as to distinguish between the effects of the H beads and
the effects of the P beads on the structure, it is helpful to study the partial
structure factors due to the two separate blocks as well as the total SSFs. 

For both the INNER--H and OUTER--H cases in conformation B (See Fig. \ref{fig:SFFInnH} and
\ref{fig:SFFOutH}), the partial SSFs of the H beads are similar to
those observed for homopolymer globules \cite{corrfn}, since the globular region
of the micelles observed in Fig. \ref{fig:phase} are densely packed. Therefore,
the total SSFs of these systems seem to correspond to denser systems than
their good solvent counterparts. Although the P beads in the INNER--H case
appear to have a similar behaviour to those of homopolymer stars in 
the good solvent
(in fact the slopes of the increase seem somewhat greater, suggesting a
stiffening of the chains) as can be seen in Figs. \ref{fig:SFFInnH}, the P
beads in the OUTER--H case behave as if they are denser, as illustrated in
Figs. \ref{fig:SFFOutH}. The partial SSF $\hat{S}_P$ appears to decrease
slower at large values of $\hat{q}$ suggesting smaller values
of the swelling exponent (inverse fractal dimension) of the P-arms
in the OUTER-H stars than the good solvent values $\nu\simeq 3/5$,
possibly quite close to the quasi--ideal coil $\nu \simeq 1/2$
as indicated by nearly constant asymptote.
This could be explained by the more compact nature of
the OUTER--H stars, due to the extra restriction on connectivity due to
attachment to the core bead as previously discussed.


\section{Conclusions}

In this paper, we have analysed the conformations adopted by a variety of
different amphiphilic star diblock copolymers involving hydrophobic (H)
and polar (P) monomers.
In particular, we have studied
the differences between stars where the hydrophobic blocks are attached to the
core monomer of the star,  INNER--H, and star where the polar blocks are
attached to the core monomer,  OUTER--H stars. Four different types of
conformations were observed. 

INNER--H stars all formed micellar--type conformations, where the hydrophobic
monomers collapsed into a globule while the polar blocks adopted coil--like
behaviour in the `shell' surrounding the globule. OUTER--H stars, on the other
hand were able to adopt three different types of conformations, depending on the
relative ratio of the lengths of the two blocks in the arms, as well as the
total length of the arms. Most OUTER--H stars were able to form
`loopy micelle' conformations, where the hydrophobic monomers collapsed
into a globule as for INNER--H topologies, however, the polar blocks in these
stars were more constrained, since they were joined at two ends, i.e., the
hydrophobic globule and the core monomer.
This extra constraint existed only in part of the phase diagram, since reducing
the relative length of the hydrophobic block, resulted in an alternative type of
conformation, where the hydrophobic blocks collapsed into small globules at the
ends of each arm, but these small globules were unable to coalesce into the
single larger globule of the loopy micellar conformations. For
intermediate values of the relative length of the hydrophobic blocks, 
the energy when these small globules coalesced 
was similar to the energy where they
did not coalesce, and hence a fourth type of conformation was observed
with different types of clustering occurring.
In this final type, the globules were able to separate and reform throughout
equilibrium simulation. This is due to that the energy barriers between the 
free energy minima corresponding to different conformations in
this regime appeared to be quite low, since relatively large specific heat
capacities were observed for these systems indicating a larger energy variance.

Increasing the length of the arms reduced the effects of the constraint of the
polar blocks in OUTER--H stars with loopy micellar conformations, and
hence these conformations were observed for a wider range of hydrophobic:polar
ratios for stars with longer arms. However, changing the number of arms did not
seem to affect the range where these conformations were observed, although
stars with more than twelve arms were not considered due to computational
limitations.

When the OUTER--H stars were able to form loopy micellar
conformations, they were more spherical and less prolate than their INNER--H
counterparts. This was due to the fact that the OUTER--H conformations are more
compact as the polar blocks are each constrained at two points.

For the two types of star topologies
we have also analysed the radial distribution functions (RDF) for the mixed
H-P  monomer pairs, which exhibited a nontrivial behaviour, in some ways 
combining features of the RDFs of P-P and H-H pairs of monomers, and
investigated the influence of the number of arms.

Finally, we have computed the total and partial H- and P- static structure
factors (SSF) of both INNER-H star micellar and OUTER-H star 
loopy micellar conformations, which could be directly tested against
experimental data. These have showed the dense structure of the
hydrophobic subglobule and nontrivial fractal dimension of
the constrained polar blocks in the loopy micelle.

Obviously these single--chain simulations do not provide
information about the inter--chain interactions which would permit one to 
investigate the issue of whether or not monomolecular micelles 
are the solitary structures for certain star copolymers
at intermediate concentrations.
However, the knowledge of single chain conformations structure 
allows us to make some predictions as to how much such unimolecular
micellar objects would interact at higher concentrations.

According to F\"orster {\it et al} \cite{ForsterMicelle1,ForsterMicelleRev}, the
mean aggregation number, Z (i.e., the number of molecules in a single micelle
above the CMC),
in diblock polymeric micelles (as well as low molecular weight surfactant
micelles) can be determined from the 
empirically determined relation: 
$Z=Z_0 N_A^2 N_B^{-0.8}$,
where $Z_0$ is a parameter which depends on the polymers or surfactants used,
and $N_A$ and $N_B$ are the degrees of polymerisation of the insoluble (H in
this case) and the soluble (P in this case) blocks, respectively. They found
that the aggregation number of triblock BAB (or PHP) copolymers could 
be determined by treating the triblock as being `cut' into two diblock halves. 
In this way the
aggregation number of the triblock is half of that for each of the two diblock
halves. They generalised this model to predict the micellesation behaviour of
graft copolymers, where a polymer with $n$ grafts is cut into $n$ pieces, each
piece containing one graft. If a constant graft density, $n/N_A$, was assumed,
the aggregation number was then predicted to decrease as
$n^{-1}$ \cite{ForsterMicelle1}.  Note, however, that a similar argument
is not valid for the ABA (or HPH) triblocks with hydrophobic ends
(end--stickers), which are known to form extended networks and telechelic
gels. 

One could also apply this principle to star diblocks, hence expecting 
that the aggregation number will decrease as $f^{-1}$. 
If this aggregation number equals 1, then this would correspond to a
`unimolecular micelle'. 
It can be seen that increasing the length of the P
block decreases this number, while increasing the length of the H block
increases this number as in F\"orster's Eq. 
By increasing the length of the H block, the surface area of `exposed' H
monomers is increased, promoting aggregation. 
Increasing the length of the P block, on the other hand, 
allows the polar monomers to shield the hydrophobic monomers from the solvent,
reducing the aggregation number as the P shell is repulsive. 
Intuitively, one could say that this shielding
is enhanced in the OUTER--H case, since the polar blocks are `wrapped' around
the hydrophobic core better. With all this in mind, one could suggest that 
OUTER--H stars with many arms, long P blocks and short H blocks are most likely 
to form only unimolecular micelles. Obviously the choice of monomers is also
important, as this affects $Z_0$. In the INNER-H case larger number of
arms $f$ would be required to form a sufficiently prohibitive polar shell to
suppress further aggregation. We hope to be able to investigate these
questions directly via multi--star simulations in the future.

\acknowledgments


We are grateful to Professor Giuseppe Allegra, Professor
Fabio Ganazzoli, Dr Guido Raos, Giovanni
Bellesia and Stefano Elli for interesting discussions.
We would also like to thank Adam Byrne for his efforts
during the course of his undergraduate project.



\newpage

\begin{figure}
\caption{ 
\label{fig:snapsC+D}
Typical snapshots of conformations of an OUTER--H f=12 arm star with each arm
100 monomers long. Fig.~a is that of a star with $N_P/N=0.92$ and Fig. b is 
of a star with $N_P/N=0.95$. 
Fig.~a corresponds to a star with conformation of type C and Fig.~b 
corresponds to a star
with conformation D. It can be seen 
that the H monomers at the end of each arm are able to cluster together to form
subglobules. In Fig.~a, type C conformation, some of these
globular ends are further able to cluster together, in this case forming 
three clusters. The number of such clusters ranged
from three to five during the course of the simulation.
}
\end{figure}

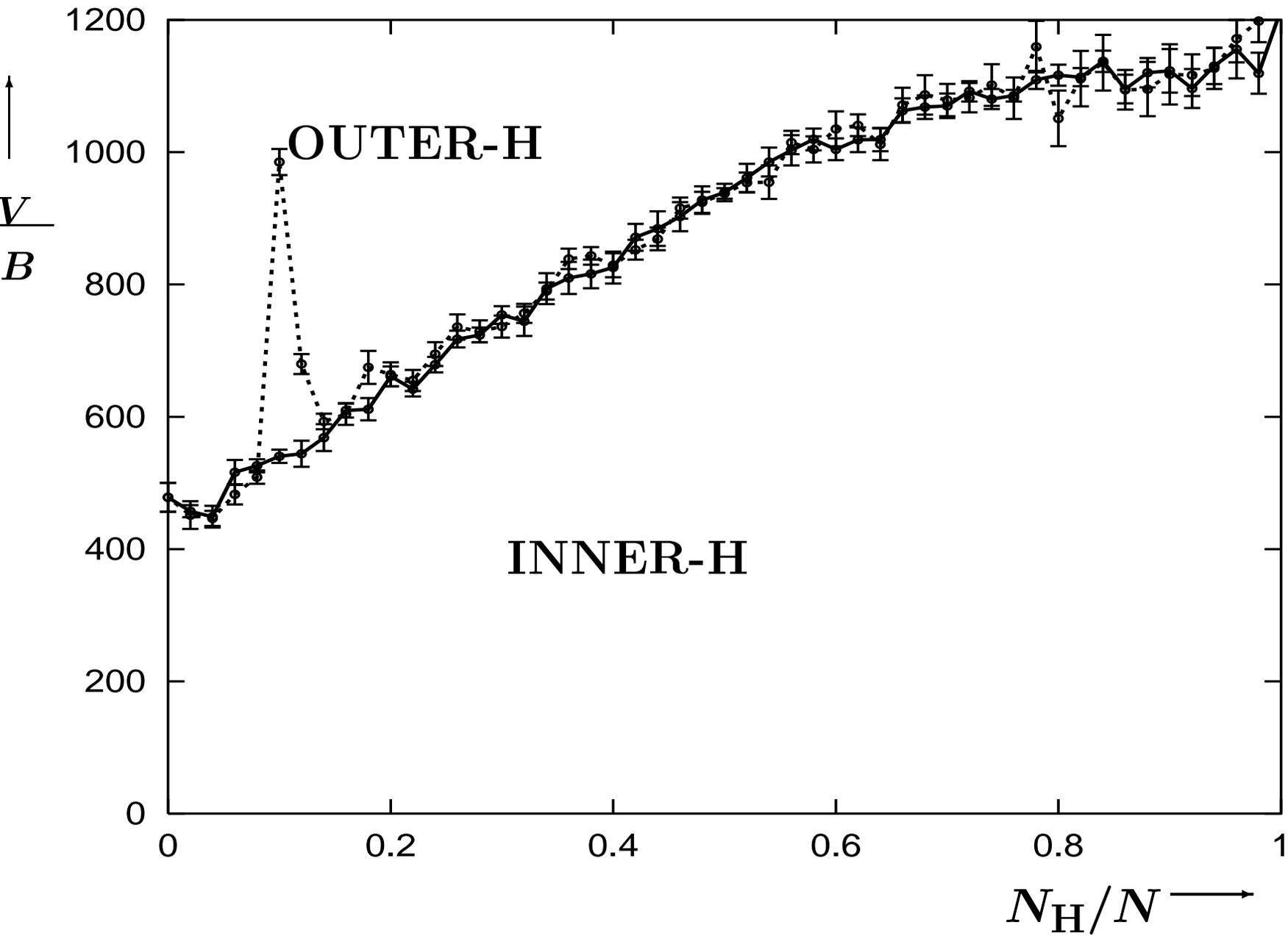
\begin{figure}
\caption{
\label{fig:C_V}
The dimensionless specific heat capacity, $c_V /k_B$, of 
INNER--H and OUTER--H stars with $f=6$ arms of $N/f=50$ monomers long 
vs the hydrophobic composition, $N_H/N$.
}
\end{figure}

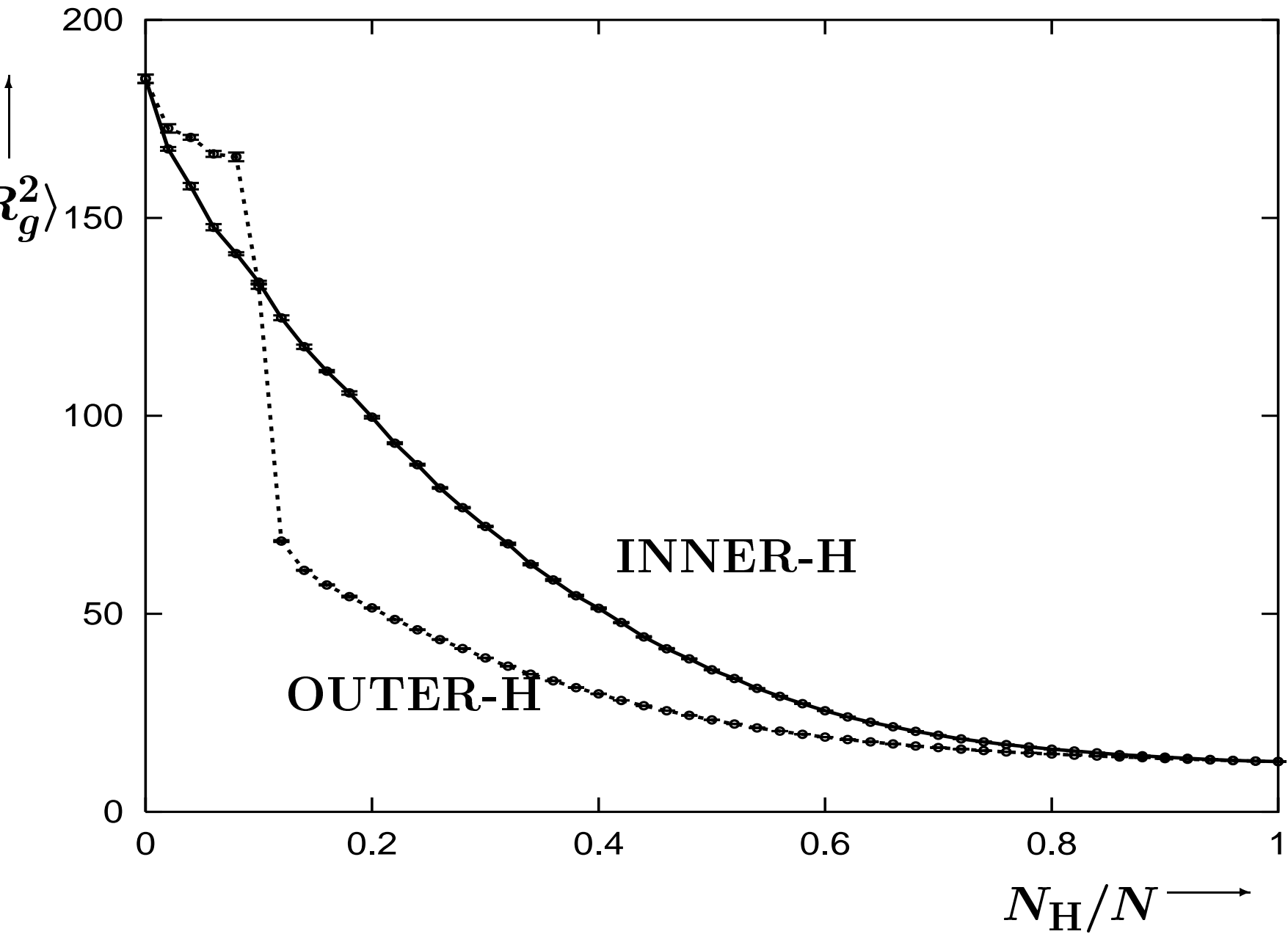
\begin{figure}
\caption{
\label{fig:R_g^2}
The mean--squared radii of gyration, $\langle R_g^2 \rangle$, 
(in units of the bond length $\ell$) of 
INNER--H and OUTER--H star
copolymers with $f=6$ arms of $N/f=50$ monomers long vs the hydrophobic
composition, $N_H/N$.  
}
\end{figure}

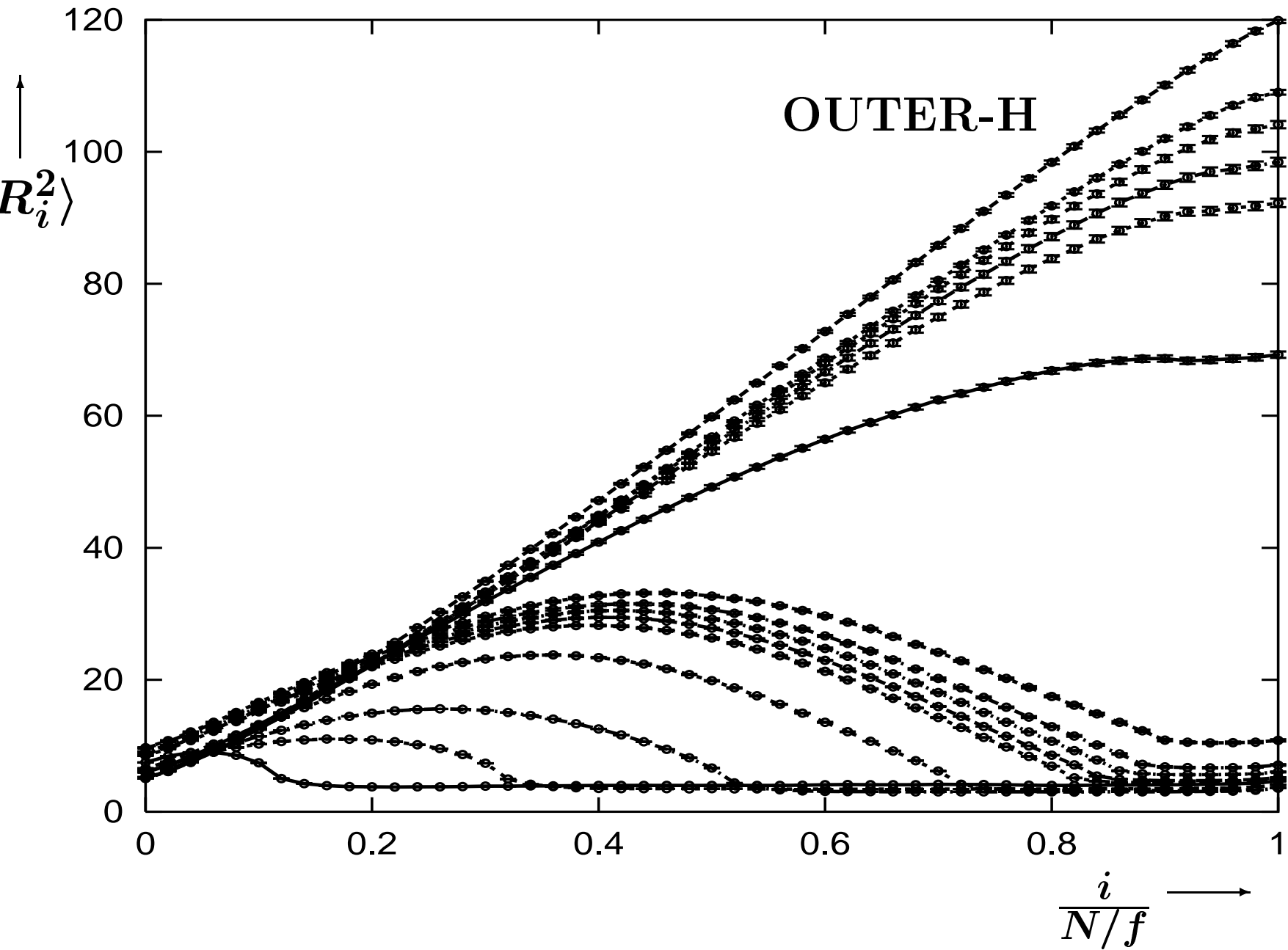
\begin{figure}
\caption{ 
\label{fig:R_i^2}
The mean--squared distances of the beads from the centre--of--mass, 
$\left\langle R_i^2 \right \rangle$, (in units of the bond length $\ell$) 
vs the chain index $i/(N/f)$ along an arm
for OUTER--H stars with $f=6$ and $N/f=50$ with
different composition ratios. Curves correspond (from bottom to top) to:
$N_{{\rm P}}/N = 0.10$, $0.30$, $0.50$, $0.70$, $0.80$, $0.82$, $0.84$,
$0.86$, $0.88$, $0.90$, $0.92$, $0.94$, $0.96$, $0.98$ and $1.00$.
}
\end{figure}

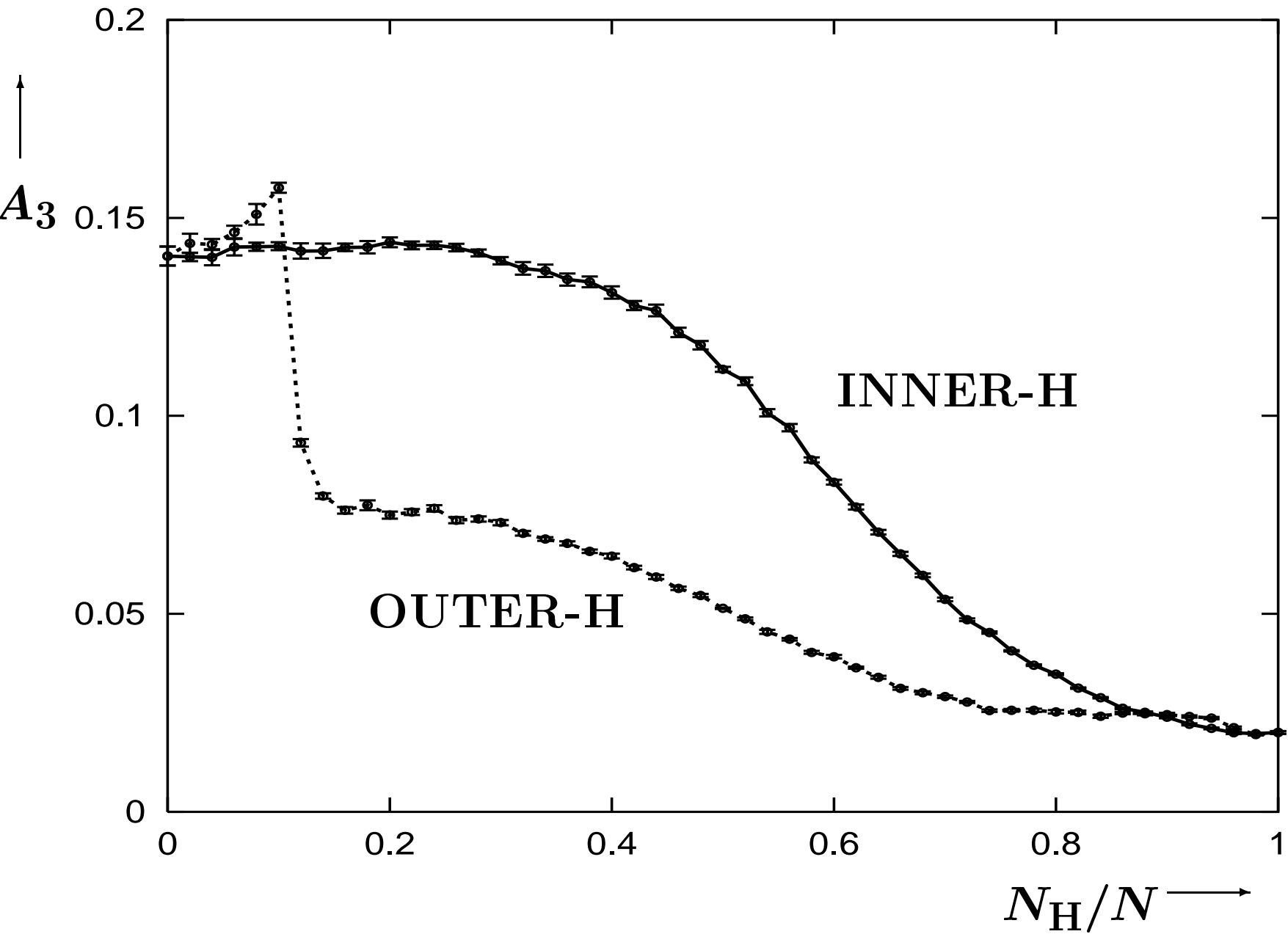
\begin{figure}
\caption{
\label{fig:A_3}
The asphericity index, $A_3$, vs the hydrophobic composition, $N_H/N$,
for INNER--H and OUTER--H stars of $f=6$ arms $N/f=50$ monomers long.
}
\end{figure}

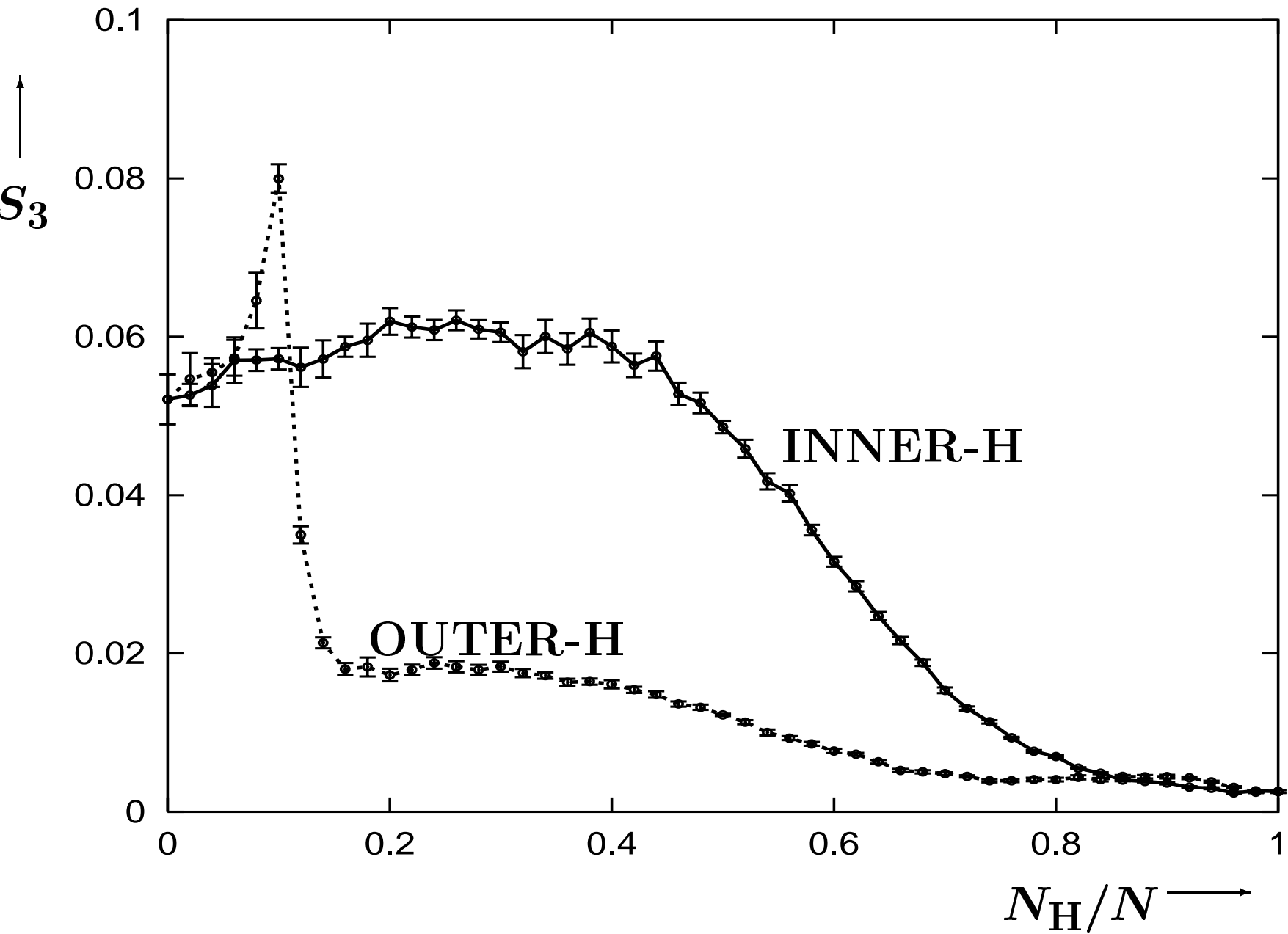
\begin{figure}
\caption{
\label{fig:S_3}
The ellipsoid index, $S_3$, vs the hydrophobic composition, $N_H/N$,
for INNER--H and OUTER--H stars of $f=6$ arms $N/f=50$ monomers long.
}
\end{figure}

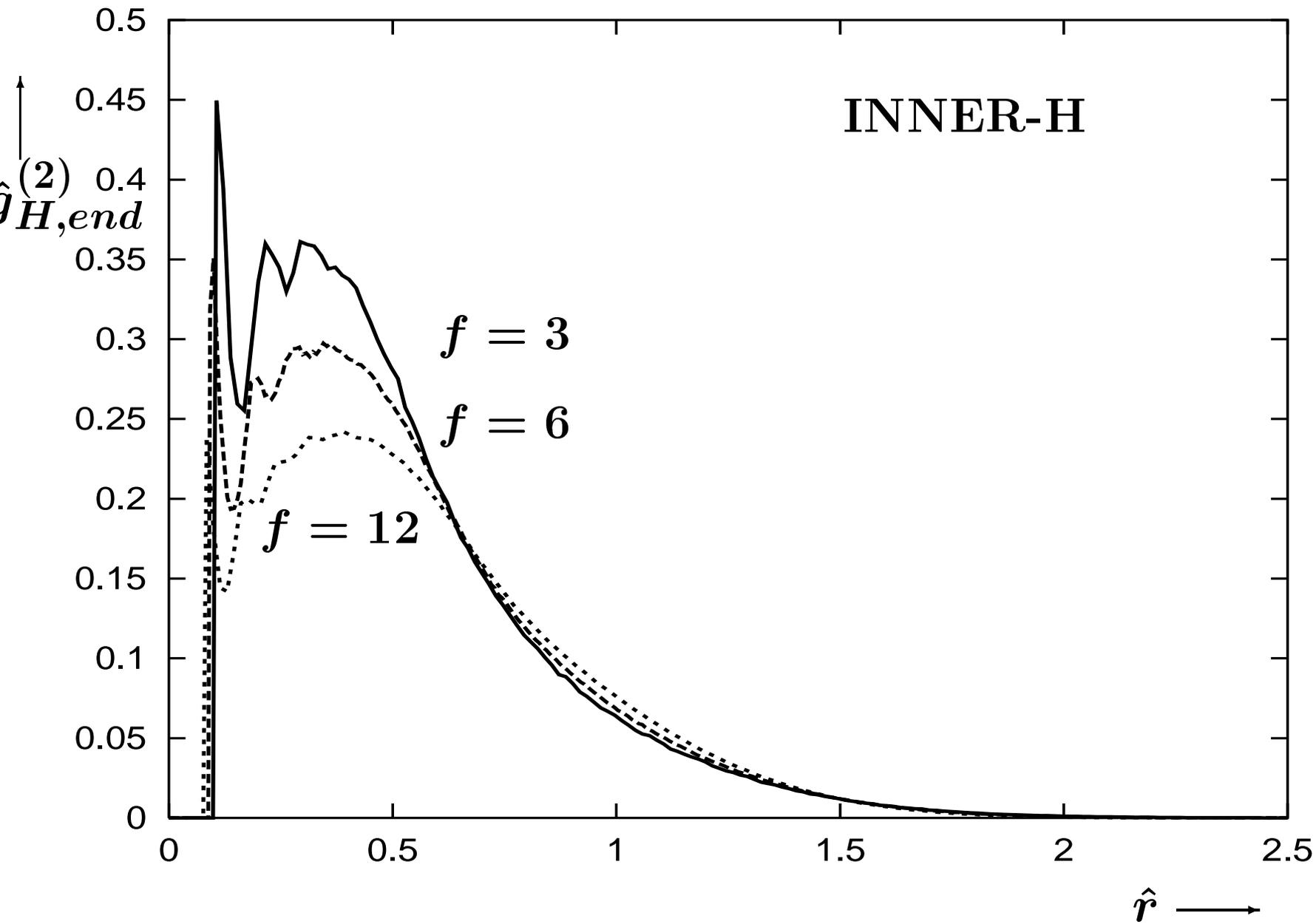
\begin{figure}
\caption{ 
\label{fig:g_rArms}
Plots of the rescaled dimensionless monomer--monomer radial distribution 
functions between the H monomers and the
end P monomers,
$\hat{g}_{H,end}^{(2)}(\hat{r})$ vs the rescaled dimensionless
radius, $\hat{r}$
for INNER--H stars with $f=3,6,12$ arms $N/f=50$ monomers long and,
$N_H/N=0.5$ composition. 
}
\end{figure}

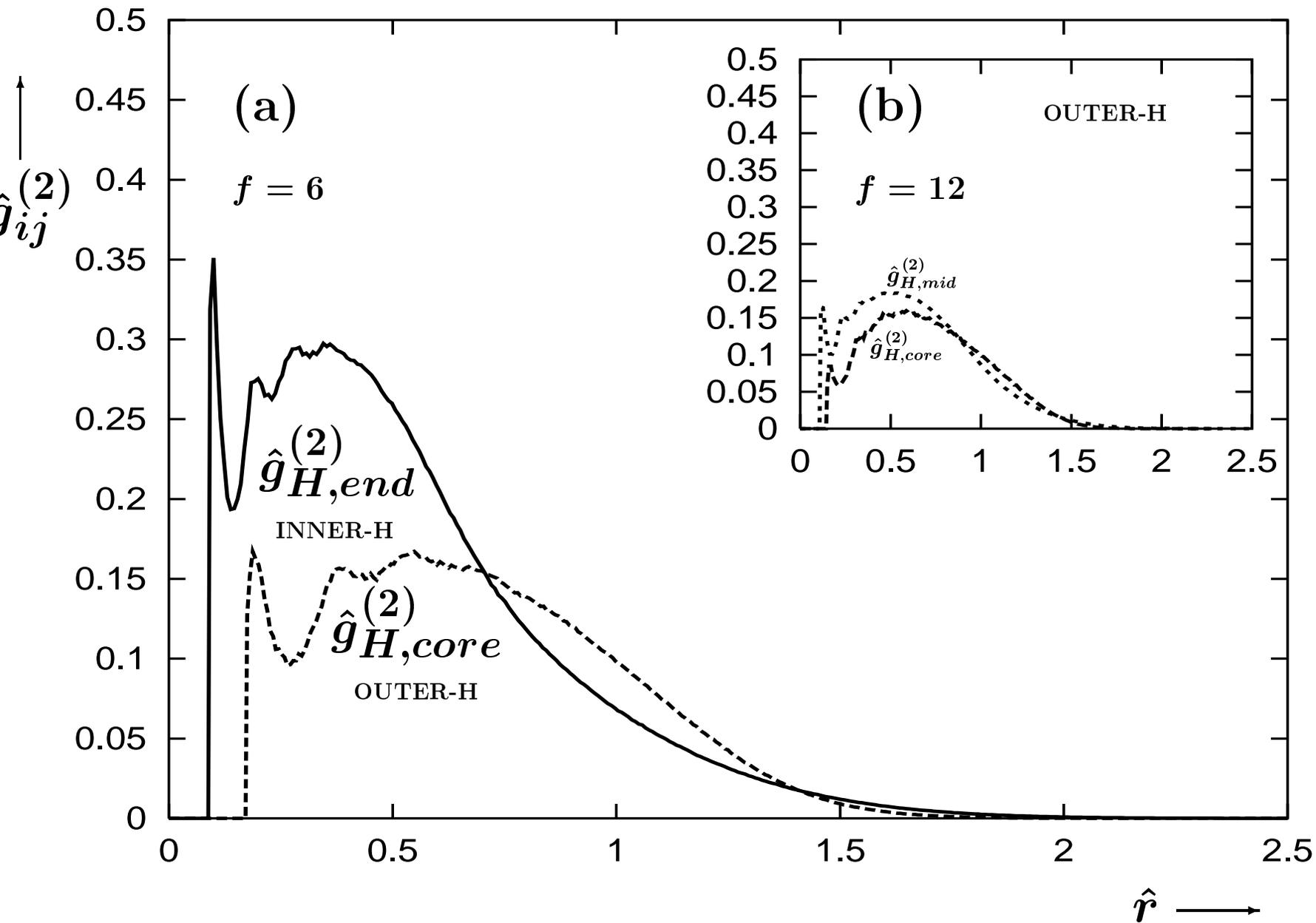
\begin{figure}
\caption{ 
\label{fig:g_rOutH}
Plots of rescaled monomer--monomer radial distribution functions,
$\hat{g}_{ij}^{(2)}(\hat{r})$ vs the rescaled radius, $\hat{r}$
for copolymers stars with $N/f=50$.
Fig.~a contains the function between the end monomers and
the H monomers of an INNER--H star (solid line) 
as well as the function between the core
monomer and the H monomers of an OUTER--H star (long--dashed line), both
of f=6 arms with $N_H/N$=0.5 composition. 
The inset (Fig.~b) contains the function between the core
monomer and the H monomers of an OUTER--H star (long--dashed line) as well as the
function between the hydrophobic monomers and the middle
monomers of the P--block (short--dashed curve) for stars with $f=12$ arms of $N/f=50$
monomers long and $N_H/N$=0.5 composition.
}
\end{figure}

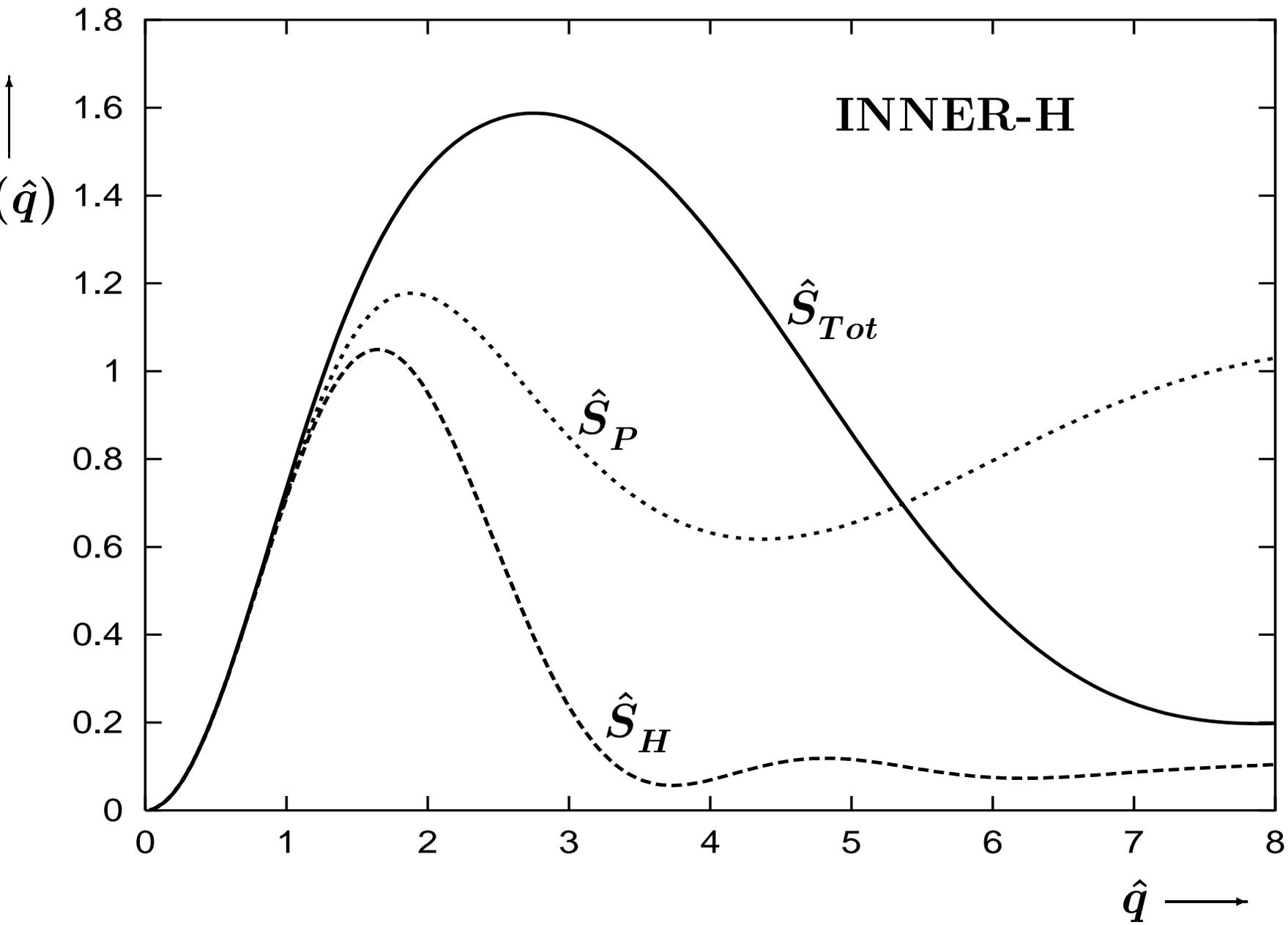
\begin{figure}
\caption{ 
\label{fig:SFFInnH}
Kratky plots of the rescaled dimensionless static structure factor 
$\hat{S}$ vs the rescaled dimensionless
wave number $\hat{q}$ for
an INNER--H star with $f=6$ arms $N/f=50$ monomers long and
$N_H/N=0.5$ composition.
Here and below the solid line corresponds to the total SSF of both H
and P beads, the long--dashed curve to the
partial SSF of the H
beads, and the short--dashed curve to the partial SSF of the P beads.
Note that the partial MS radii of gyration have been used to rescale 
the partial SSF.
}
\end{figure}

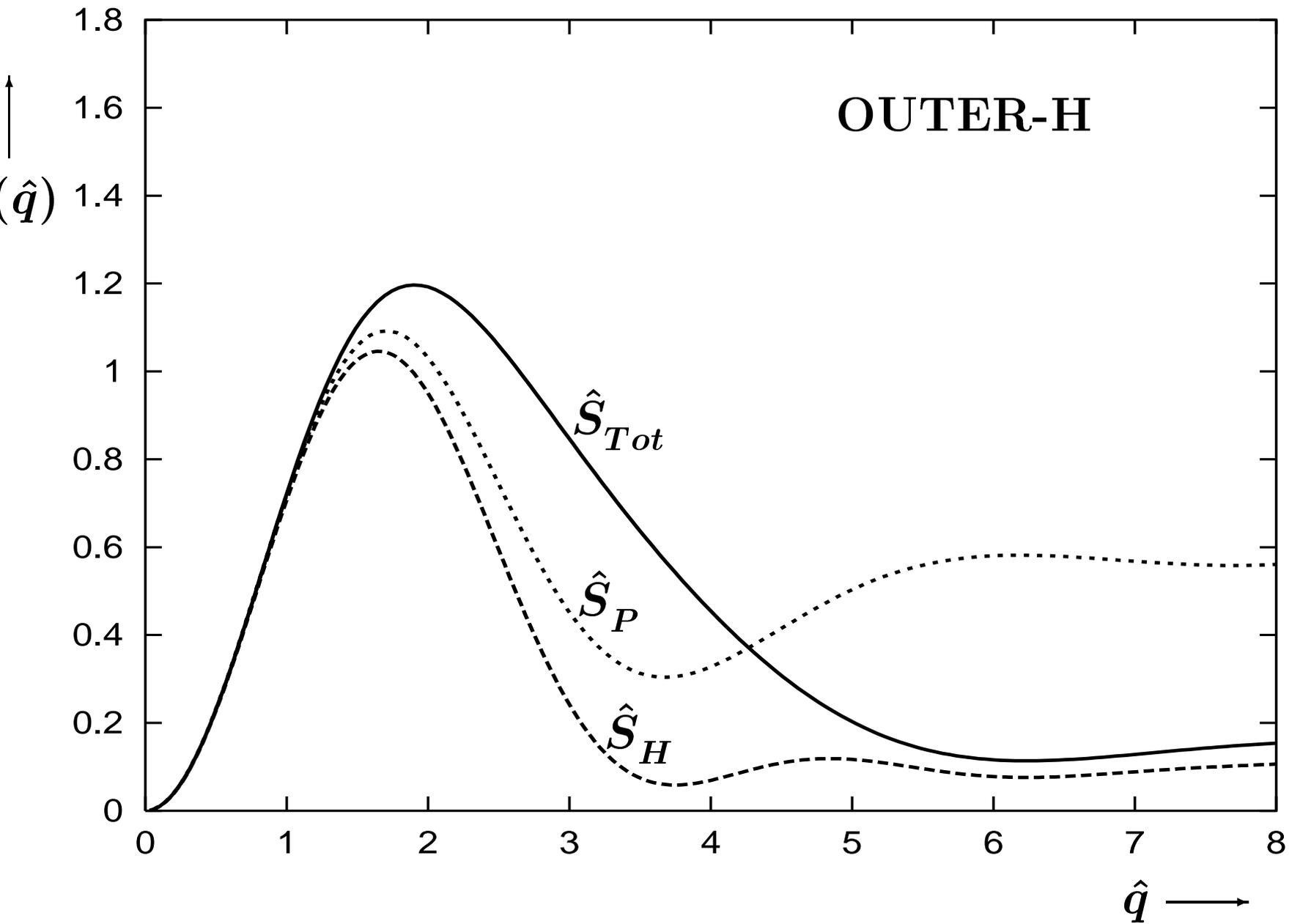
\begin{figure}
\caption{ 
\label{fig:SFFOutH}
Kratky plots of the rescaled static structure factor 
$\hat{S}$ vs the rescaled
wave number $\hat{q}$ for
an OUTER--H star in conformation type B with $f=6$ arms $N/f=50$ monomers long
and $N_H/N=0.5$ composition.
}
\end{figure}


\begin{references}
\bibitem{footnote}
Note that for the heat capacity, the following formula should be used
instead for the error: 
$\pm (N(k_B T)^2 \sqrt{Q})^{-1}\sqrt{\langle (\Delta E)^4\rangle 
-\langle (\Delta E)^2\rangle^2}$.
\end{references}
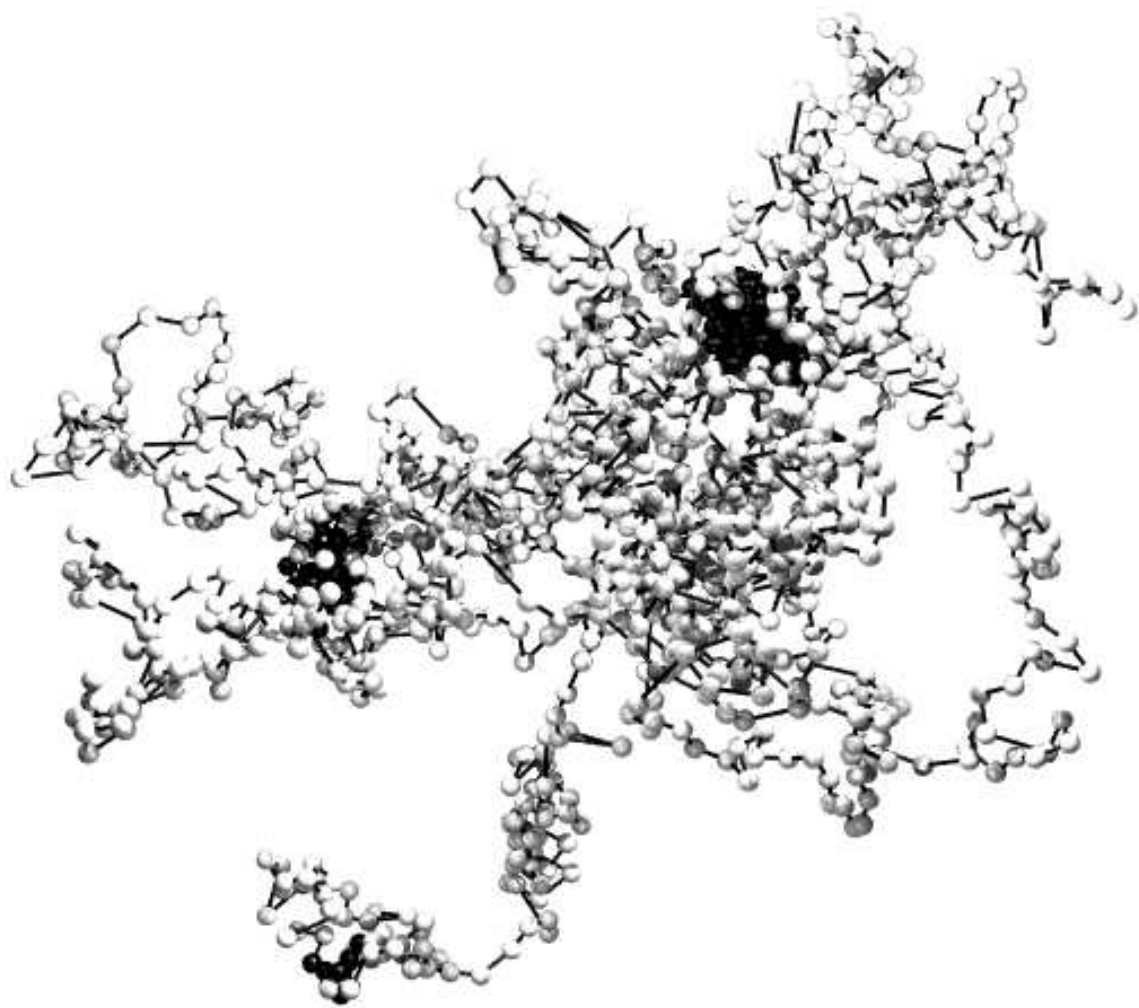
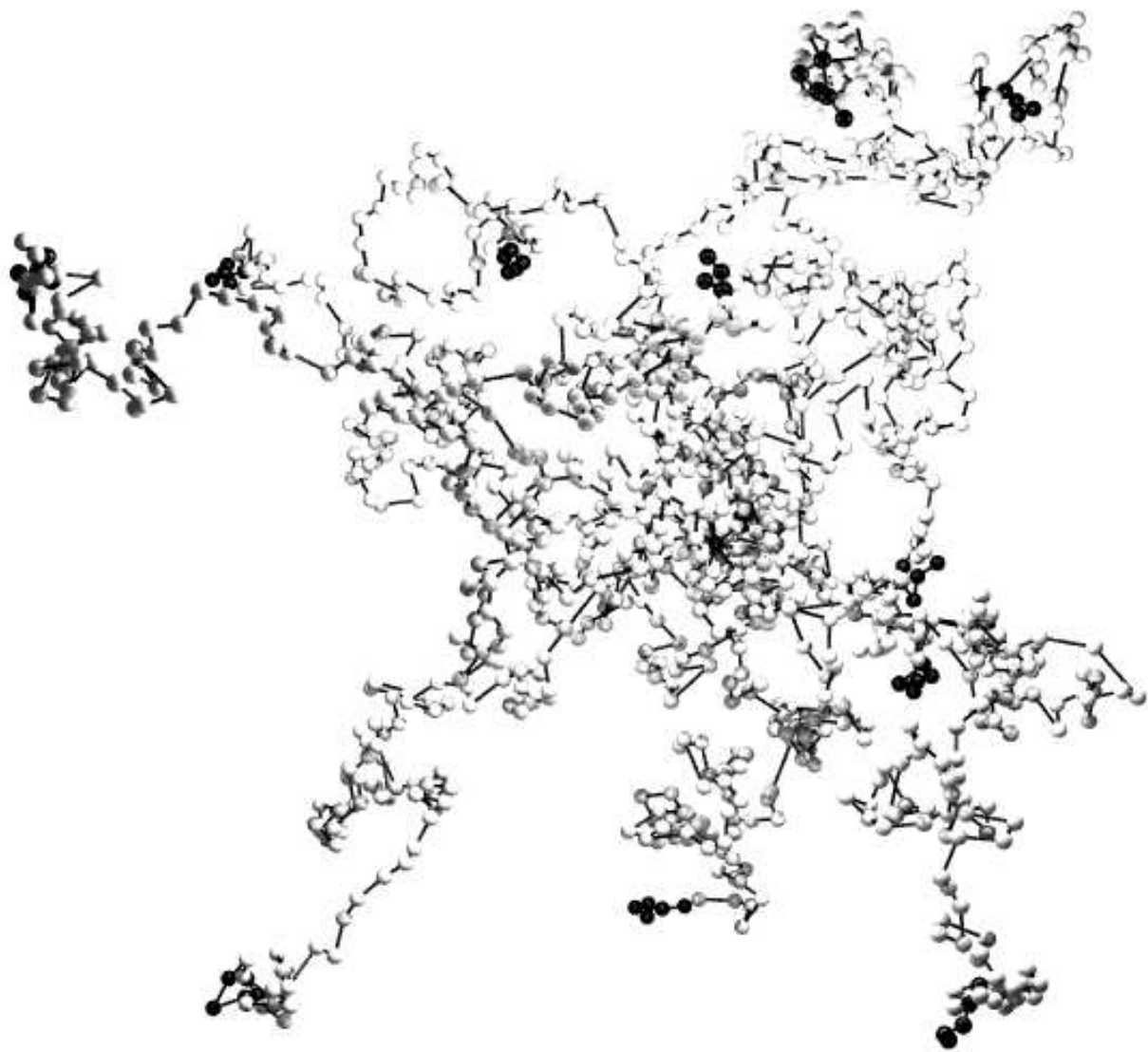
\end{document}